\documentclass[smallextended]{svjour3-arxiv}       % onecolumn (second format)
\smartqed  % flush right qed marks, e.g. at end of proof
\usepackage{graphicx}
\usepackage[top=30truemm,bottom=30truemm,left=25truemm,right=25truemm]{geometry}
\usepackage{amsmath}
\usepackage{amsfonts}
\usepackage{bm}
\usepackage{color}
\definecolor{blue}{rgb}{0, 0.4470, 0.7410}
\definecolor{red}{rgb}{0.8500, 0.1250, 0.0480} 
\definecolor{green}{rgb}{0.4660, 0.6740, 0.1880}

\newcommand{\fg}[1]{{\color{black}#1}}

\journalname{}
\begin{document}

\title{Inserting machine-learned virtual wall velocity for large-eddy simulation\\of turbulent channel flows}

\author{Naoki Moriya \and Kai Fukami \and Yusuke Nabae\\
  Masaki Morimoto \and Taichi Nakamura \and Koji Fukagata}

\institute{Naoki Moriya, Yusuke Nabae, Masaki Morimoto, Taichi Nakamura, and Koji Fukagata \at
              Department of Mechanical Engineering, Keio University, Yokohama, 223-8522, Japan \\
              \email{fukagata@mech.keio.ac.jp}
              \and
              Kai Fukami \at
              Department of Mechanical and Aerospace Engineering, University of California, Los Angeles, CA 90095, USA}

\date{\today}

\maketitle

\begin{abstract}

We propose a supervised-machine-learning-based wall model for \fg{coarse-grid wall-resolved} large-eddy simulation (LES).
Our consideration is made on LES of turbulent channel flows with a first grid point set relatively far from the wall ($\sim$ 10 wall units)\fg{, while still resolving the near-wall region,}
to present a new path to save the computational cost. 
Convolutional neural network (CNN) is utilized to estimate a virtual wall-surface velocity from $x-z$ sectional fields near the wall, whose training data are generated by a direct numerical simulation (DNS) at ${\rm Re}_{\tau}=180$.
The virtual wall-surface velocity is prepared with the extrapolation of the DNS data near the wall.
This idea enables us to give a proper wall condition to correct a velocity gradient near the wall.
The estimation ability of the model from near wall information is first investigated as {\it a priori} test. 
The estimated velocity fields by the present CNN model are in statistical agreement with the reference DNS data. 
The model trained in {\it a priori} test is then combined with the LES as {\it a posteriori} test.
We find that the LES can successfully be augmented using the present model 
at both the friction Reynolds number ${\rm Re}_{\tau}=180$ used for training and the unseen Reynolds number ${\rm Re}_{\tau}=360$ even when the first grid point is located at 5 wall units off the wall.
We also investigate the robustness of the present model for the choice of sub-grid scale model and the possibility of transfer learning in a local domain.
The observations through the paper suggest that the present model is a promising tool for recovering the accuracy of LES with a coarse grid near the wall.
\end{abstract}

\section{Introduction}

Large-eddy simulation (LES) has played a crucial role for mechanical and aerospace engineering applications in a practical manner.
Capturing momentum transfer and near wall behaviors aided by LES in a reasonable accuracy enable us to analyze complex turbulence phenomena and also understand flow physics.
However, it requires the massive computational power with the gigantic number of discretized grid points to handle these simulations since turbulence includes a wide range of scales inside them.

To avoid enormous computational cost to resolve near-wall structures, LES at practically high Reynolds numbers is often performed by modeling the flow in the near-wall region and imposing the boundary condition off the wall (i.e., wall-modeled LES).
The well-used strategy for high Reynolds number LES is to resolve turbulent structures in the outer layer region corresponding to approximately 90\% of the boundary layer directly, while modeling the rest of 10\% in the inner layer region~\cite{SJSA1997}.
The fact that the computational cost increases by more than
the square of the Reynolds number also supports the use of aforementioned strategy~\cite{SJSA1997,KT2017}. 
Hence, the utilization of a proper wall model is unavoidable to date for capturing near wall behaviors with the reasonable number of computational grid points.
Such efforts on the wall model can mainly be divided into two types: $\rm (i)$ hybrid model with RANS (e.g., detached eddy simulation, DES)~\cite{Spa2009,Ham2001} and $\rm (ii)$ methods to approximate/augment the wall-shear stress~\cite{PB2002}.
The variety of open source codes enables us to access some implementations of the former model easily in recent years, but the discontinuity between LES and RANS is known as one of the problems, which requires an artificial manipulation to avoid the inconsistency with regard to velocity there~\cite{NNWSS2000,MPMA2008}.
The latter has been carried out for a long time with the use of an artificial boundary condition. 
The sophisticated concept has recently been developed, in which viscous-scale grids are embedded in the inner layer so as to solve the boundary layer equation, and its effectiveness against the above velocity mismatch has been reported in Kawai and Larsson~\cite{KL2012}.
However, it is still difficult to accurately predict the turbulent boundary layer at high Reynolds numbers without tuning of empirical parameters and the use of complex control theories.

An open issue is present also for LES
with no-slip boundary condition (i.e., wall-resolved LES) if one wants to locate the first grid point outside the viscous sublayer to save the computational cost while resolving the near-wall structure. 
Consider, for instance, the boundary condition is discretized using a second-order central difference on a staggered grid system, 
the no-slip boundary condition for the streamwise velocity, say $u_w=0$, on the wall ($y=0$) can be discretized to satisfy $(u_0+u_1)/2=0$, i.e.,
$u_0 = -u_1$, where $u_0$ and $u_1$ denote the streamwise velocity $u$ at $y_0=-\Delta y/2$ (i.e., the first grid point outside the boundary) and $y_0=\Delta y/2$ (i.e., the first grid point off the wall), respectively, with $\Delta y$ being the size of first wall-normal grid,
while the velocity gradient on the wall is discretized as $(\partial u/\partial y)_w=(u_1-u_0)/\Delta y$. 
Namely, $u_1$ is always computed as $u_1 = u_w + (\partial u/\partial y)_w \Delta y/2$.
This is reasonable as far as the first grid point is located in the viscous sublayer where the linear law $\overline{u}^+=y^+$ (where $\overline{u}$ is the mean streamwise velocity, and the superscript ``+'' denotes the wall units) holds.
However, this treatment becomes inappropriate when the first grid point is farther off the wall.
Since the stress balance near the wall ($y^+\ll {\rm Re}_\tau$, where ${\rm Re}_\tau$ denotes the friction Reynolds number) can be expressed in wall units as
\begin{equation}
\frac{\partial \overline{u}^+}{\partial y^+} - \overline{u^{\prime +} v^{\prime +}} = 1,
\end{equation}
imposition of no-slip boundary condition in the original form overestimates the velocity gradient on the wall by the amount of Reynolds shear stress $- \overline{u^{\prime +} v^{\prime +}}$ when the first grid point is located outside the viscous sublayer --- and this why it is usually advised to place several grid points in the viscous sublayer for a wall-resolved LES. 
Thus, a proper correction is required for the imposition of discretized no-slip boundary condition
even if the Reynolds shear stress is perfectly amended by the SGS model.
This argument is similar to the discussion by Kawai and Larsson~\cite{KL2012} on the log-layer mismatch; however, for wall-resolved LES, similar corrections should be required not only for the mean streamwise velocity but also for all the fluctuating velocity components to capture the near-wall coherent structures.

To address the aforementioned issues, machine learning, which has been known as a good candidate to handle complex fluid flow problems~\cite{BNK2020,BEF2019}, can be an attractive tool. 
As for the application to closure modeling, the machine learning has already had a citizenship there~\cite{DIX2019,DK2021}.
One of the seminal works is tensor-basis neural network (TBNN) with Galilean invariance embedded by Ling et al.~\cite{LKT2016}.
Their model was tested for duct and wavy-wall flows.
We have recently been able to see the extension of TBNN to various flow configurations and problem settings, e.g., channel flow at various Reynolds numbers~\cite{zhang2019application}, a cylindrical and inclined jet in crossflow~\cite{milani2021turbulent}, and the pressure-Hessian based closure~\cite{parashar2020modeling}.
For the application to LES, the idea to estimate finer (unresolved) scales from solved large-scale information has widely been accepted with the supervised machine learning, whose training data is prepared by direct numerical simulation (DNS)~\cite{gamahara2017searching,maulik2017neural,maulik2019subgrid,maulik2018data,maulik2019sub,beck2019deep,yang2019predictive,pawar2020priori,KKWL2021}.

In the present study, we focus on the capability of machine learning for data estimation and reconstruction towards the augmentation of LES, rather than the closure modeling efforts.
Because the machine learning is good at extracting hidden features of data, it has also been widely utilized for state estimation tasks~\cite{FFT2019,FFT2020b}.
Guastoni et al.~\cite{guastoni2020convolutional} reported that a convolutional neural network (CNN) is able to estimate the state of turbulent channel flow from only wall-sensor measurements by combining to proper orthogonal decomposition.
Toward the combination with the opposition control, there are several studies that aim at estimating the velocity field at $y^+\approx 15$ in a turbulent channel flow from the wall measurements~\cite{han2020active,park2020machine}.
More recently, Nakamura et al.~\cite{nakamura2021comparison} compared the capability of linear methods and neural networks in state estimation of minimal turbulent channel flow.
These results suggest the hidden relation between the state of turbulent flows and wall quantities.
In turn, this also motivates us to expect the possibility to estimate the wall information from the state above the wall.
Obtaining the clues from these pieces above, we propose a machine-learning-based wall model
%ing 
for \fg{wall-resolved} LES.
% of wall turbulence.
Especially, we focus on giving a proper wall condition which corrects a velocity gradient near the wall for the case with a very coarse staggered grid,
\fg{with the first point from the wall being located at $y^+\sim 10$.}
% in the wall-normal direction.
Our model aims to insert the machine-learning-based artificial slip velocity for corrections of not only the mean velocity profile but also all the fluctuating velocity components.

The present paper is organized as follows: we introduce the overview with the covered regression methods in Section~\ref{sec:2}.
The construction of the present wall model ({\it a priori} test) and its application to the LES ({\it a posteriori} test) are expressed in Section~\ref{sec:3}.
Concluding remarks are provided in Section~\ref{sec:4}.

% \vspace{-4mm}
\section{Methods}
\label{sec:2}

\subsection{Overview of the present wall modeling for large-eddy simulation}
\label{sec:2-1}

%%%%%%%%%%%%%%%%%%%%%%%%
%%figure 1%%%%%%%%%%%%%%
%%%%%%%%%%%%%%%%%%%%%%%%
\begin{figure}
  \centering
  \includegraphics[width=1.\textwidth]{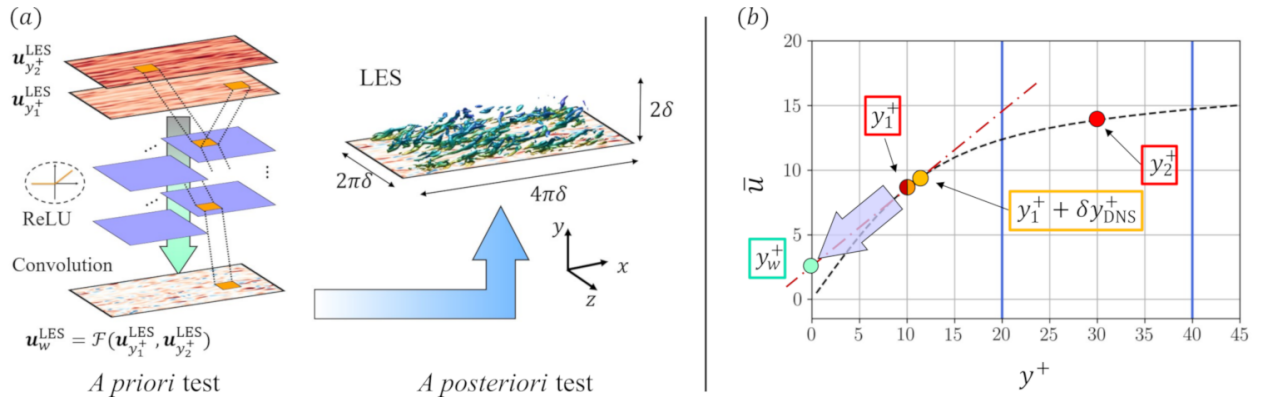}
  \caption{The present machine-learning-based wall modeling in LES. $(a)$ Concept of {\it{a priori}} test and {\it{a posteriori}} test. $(b)$ Preparation for a virtual wall surface velocity ${\bm u}_{y_w}$. The vertical blue lines at $y^+=0$, $20$ and $40$ represent the locations of cell faces in LES on a staggered grid.}
  \label{fig1}
\end{figure}
%%%%%%%%%%%%%%%%%%%%%%%%
%%%%%%%%%%%%%%%%%%%%%%%

The concept of this study is mainly composed of two parts --- {\it{a priori}} test and {\it{a posteriori}} test --- as illustrated in Fig.~\ref{fig1}.
As described in introduction part, we aim at reducing the number of grid points in the near-wall region in an LES using a staggered grid system as much as possible, while still resolving the near-wall flow structure.
However, when we apply a no-slip wall boundary condition in the wall-normal direction, a velocity at the first point from the wall may be overestimated as mentioned in the introduction, which causes non-negligible error in the mean velocity distribution.
Similarly, although not illustrated in Fig.~\ref{fig1}, all the fluctuating velocity components are also subjected to the similar discretization error.
Therefore, we here consider the artificial wall slip velocity ${\bm u}_w$ to fix all the velocity components at the first point from the wall ${\bm u}_{y_1}$.

In {\it{a priori}} test, a machine-learning model ${\cal F}$ is constructed to estimate the artificial slip velocity ${\bm u}_w$, from the velocity on two $x-z$ cross sections near wall region $\{{\bm u}_{y_1},{\bm u}_{y_2}\}$
such that ${\bm u}_w \approx {\cal F}({\bm u}_{y_1},{\bm u}_{y_2})$, where $y_1$ and $y_2$ denote the locations of the first and second grid points used in LES.
Hereafter, this artificial wall velocity is referred to as {\it virtual wall surface velocity}. 
For the training of machine-learning model, we use subsampled $x-z$ cross-sectional velocity data $\tilde{\bm u}$ obtained by a direct numerical simulation (DNS) at the same friction Reynolds number ${\rm Re}_\tau=180$ as that used in the target LES.
The subsampling operation for the DNS data enables us to match the streamwise and spanwise grid resolutions to those in the target LES.
As illustrated in Fig.~\ref{fig1}$(b)$, the virtual wall surface velocity ${\bm u}_w$ used for a training process is prepared using the linear extrapolation from the DNS data at the first point from wall in the target LES, $y^+_1$ (where ``+'' denotes the wall units), and the next point in DNS, $y^+_1+\delta y^+$, as
\begin{equation}
\tilde{\bm{u}}^{\rm{DNS}}_{w}=
\tilde{\bm{u}}^{\rm{DNS}}_{y^+_1}
-\frac{
\tilde{\bm{u}}^{\rm{DNS}}_{y^+_1+\delta y^+}
-\tilde{\bm{u}}^{\rm{DNS}}_{y^+_1}
}
{\delta y^+}
y^+_1.  
\end{equation}
Here, the subsampling operation is denoted as $\tilde{(\cdot)}$.
In {\it{a priori}} test, a machine-learning model $\cal F$ estimates the virtual wall surface velocity $\tilde{\bm{u}}_{w}^{\rm{ML}}$ from the two cross-sectional DNS data at the first and the second points from the wall in a target LES, $\tilde{\bm{u}}_{y^+_1}^{\rm DNS}$ and $\tilde{\bm{u}}_{y^+_2}^{\rm DNS}$, as
\begin{equation}
    \tilde{\bm{u}}_{w}^{\rm{ML}}={\cal{F}}(\tilde{\bm{u}}_{y^+_1}^{\rm DNS}, \tilde{\bm{u}}_{y^+_2}^{\rm DNS}),
\end{equation}
and it is compared with the reference, $\tilde{\bm{u}}_{w}^{\rm{DNS}}$. 
The constructed model ${\cal F}$ is then applied to an LES in {\it{a posteriori}} test by using it as the boundary condition, i.e.,
\begin{equation}
    \bm{u}_{w}^{\rm{LES}}=\tilde{\bm{u}}_{w}^{\rm{ML}}={\cal{F}}({\bm{u}}_{y^+_1}^{\rm LES}, {\bm{u}}_{y^+_2}^{\rm LES}).
 \end{equation}

\subsection{Convolutional neural network}

%%%%%%%%%%%%%%%%%%%%%%%%
%%figure 2%%%%%%%%%%%%%%
%%%%%%%%%%%%%%%%%%%%%%%%
\begin{figure}
  \centering
  \includegraphics[width=0.97\textwidth]{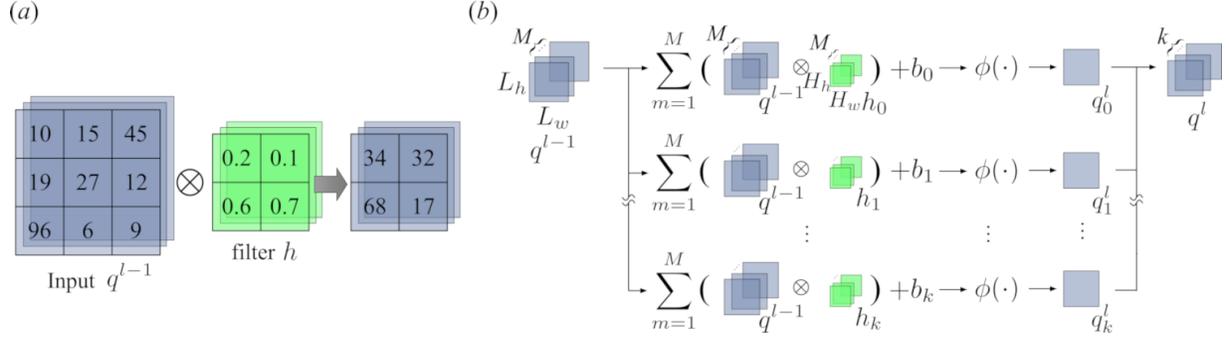}
  \caption{Internal operations of convolutional neural network: convolutional operations $(a)$ at each channel and $(b)$ at layer.}
  \label{Conv_opr}
\end{figure}
%%%%%%%%%%%%%%%%%%%%%%%%
%%%%%%%%%%%%%%%%%%%%%%%

As a machine-learning model, we capitalize on convolutional neural network (CNN)~\cite{LBBH1998} originally developed in image recognition.
The filters, trainable parameters inside the CNN, are able to handle high-dimensional data efficiently and extract key features.
Thanks to its unique capability in handling high-dimensional data, the use of CNN has also been spread in the fluid dynamics field in recent years~\cite{FFT2020,FNKF2019,SP2019,MFF2019,FNF2020,FHNMF2020,fukami2020sparse,morimoto2021convolutional,FukamiVoronoi,matsuo2021supervised}.

As shown in Fig.~\ref{Conv_opr}$(a)$, the basic operation of CNN is to take a summation of a Hadamard product between a designated region of input data and a trainable filter $h$.
Usually, a CNN consists of several convolutional layers to build a certain relationship between inputs and outputs.
In this study, velocity fields of $x-z$ cross-sections at two designated $y^+$ are fed into the first convolutional layer and then $q^{(1)}$ will be obtained as an output of the first layer.
The procedure in obtaining $q^{(l)}$ from $q^{(l-1)}$ is repeated until $l<l_{\rm max}$, where the final output $q^{(l_{\rm max})}$ corresponds to a virtual wall surface velocity in our case.
The operation inside the convolutional layer can be expressed as,
\begin{equation}
    q^{(l)}_{ijk}=\phi\left(\sum_{m=1}^M\sum_{p=0}^{H_h-1}\sum_{q=0}^{H_w-1}h^{(l)}_{pqmk}q^{(l-1)}_{i+p-C,j+q-C,m}+b_k^{(l)}\right),
\end{equation}
where $C={\rm floor}(H/2)$, $b_k^{(l)}$ is a bias, $\phi$ is an activation function, $M$ is the number of input data channels and $k$ is the number of filters (equals to number of output data channels), respectively. 

\begin{table}[tbp]
    \centering
    \vspace{3mm}
    \caption{The structure of the machine learning model. The convolution layer is denoted as Conv2D.}
    \vspace{3mm}
    \label{Structure_model}
    \begin{tabular}{ccc}\hline\hline
         Layer (filter size, \# of filters) & Data size & Activation\\ \hline
         Input & (64,64,2)& \\
         1st Conv2D (5,32) & (64,64,32) & ReLU\\
         2nd Conv2D (5,32) & (64,64,32) & ReLU\\
         3rd Conv2D (5,32) & (64,64,32) & ReLU\\
         4th Conv2D (5,32) & (64,64,32) & ReLU\\
         5th Conv2D (5,32) & (64,64,32) & ReLU\\
         6th Conv2D (5,1)  & (64,64,1)  & Linear\\ \hline\hline
    \end{tabular}
\end{table}

The details of the proposed model are summarsized in Table~\ref{Structure_model}.
The weights on the filters ${\bm w}$ are optimized through the back propagation \cite{Kingma2014} by minimizing a cost function computed from output ${\bm q}^{(l_{\rm max})}$ and reference data ${\bm q}_{\rm ref}$, such that
\begin{align}
        {\bm w}&={\rm argmin}_{\bm w}||{\bm q}^{(l_{\rm max})}-{\bm q}_{\rm r
    ef}||_2\nonumber\\ 
    &={\rm argmin}_{\bm w}||{\cal F}(\{\tilde{\bm u}^{\rm DNS}_{y_1^+},\tilde{\bm u}^{\rm DNS}_{y_2^+}\};{\bm w})-\tilde{\bm u}^{\rm DNS}_w||_2. 
    \label{eq2}
\end{align}
We use the $L_2$ error as the cost function.

\subsection{Linear regression analysis}

To clarify the advantage of nonlinear CNN, we also perform a linear regression analysis for virtual wall-surface velocity estimation~\cite{nakamura2021comparison,SH2017}.
The linear regression is able to express the output as a linear map of input,
\begin{equation}
    {\bm Q} = {\bm P}{\bm \beta},
\end{equation}
where ${\bm P}$, ${\bm Q}$, ${\bm \beta}$ is an input matrix, an output matrix, and an weight matrix, respectively.
Since we use two velocity sectional fields $\{\tilde{\bm u}^{\rm DNS}_{y_1^+}$, $\tilde{\bm u}^{\rm DNS}_{y_2^+}\}$ as the input data, the virtual wall surface velocity estimated by the linear regression $\tilde{\bm u}^{\rm LR}_{w}$ is expressed as the linear sum of input velocity and weight, such that
\begin{equation}
\tilde{\bm u}^{\rm LR}_{w} = \tilde{\bm u}^{\rm DNS}_{y_1^+} {\bm \beta_1} + \tilde{\bm u}^{\rm DNS}_{y_2^+} {\bm \beta_2},
    \label{eq1}
\end{equation}
where a single snapshot of velocity data is reshaped into a one-dimensional vector, ${\bm \beta_1}$ and ${\bm \beta_2}$ are the weight matrices. 
The weight matrices ${\bm \beta_1}$ and ${\bm \beta_2}$ are optimized so that the $L_2$ norm between the left-hand side and the right-hand side of Eq.~\ref{eq1} can be minimized over the trained snapshots. 
It can mathematically be formed as
\begin{equation}
{\bm \beta}_1, {\bm \beta}_2 = {\rm argmin}_{\beta_1, \beta_2}
\|\tilde{\bm u}^{\rm DNS}_{w} - (\tilde{\bm u}^{\rm DNS}_{y_1^+} {\bm \beta_1} + \tilde{\bm u}^{\rm DNS}_{y_2^+}{\bm \beta_2})\|_2.
\end{equation}
We do not use the $L_1$ or $L_2$ penalization terms for the fair comparison to the CNN (i.e., Eq.~\ref{eq2})~\cite{nakamura2021comparison}.

%%%%%%%%%%%%%%%%%%%%%%%%%%%%%%%%%%%%%%%%%
%             table 1
%%%%%%%%%%%%%%%%%%%%%%%%%%%%%%%%%%%%%%%%%
\begin{table*}
    \vspace{3mm}
    \centering
    \vspace{3mm}
    \caption{The covered grid widths in the $y$ direction.}
    \label{grid_width}
    \begin{tabular}{cccc}\hline
    \hline
        \\[-7pt]
        $(y^+_1, y^+_2)$ & $y^+_1+\delta y^+_{\rm DNS}$ & $\Delta y^+$ & $N^{\dag}_y$\\
        \\[-7pt]
        \hline
        \\[-7pt]
        %\hline
        $(2.50, 7.50)$ &  3.75 & $5.00$ & 72\\
        %\hline
        $(5.00, 15.0)$ & 6.25 & $10.0$ & 36\\
        %\hline
        $(10.0, 30.0)$ & 11.3 & $20.0$ & 18\\
        \hline
        \hline
    \end{tabular}
\end{table*}
%%%%%%%%%%%%%%%%%%%%%%%%%%%%%%%%%%%%%%%%%
%             table 1
%%%%%%%%%%%%%%%%%%%%%%%%%%%%%%%%%%%%%%%%%

% \hspace{5px}
%%%%%%%%%%%%%%
%  figure   %
%%%%%%%%%%%%%%
\begin{figure}
  \centering
  \includegraphics[width=0.95\textwidth]{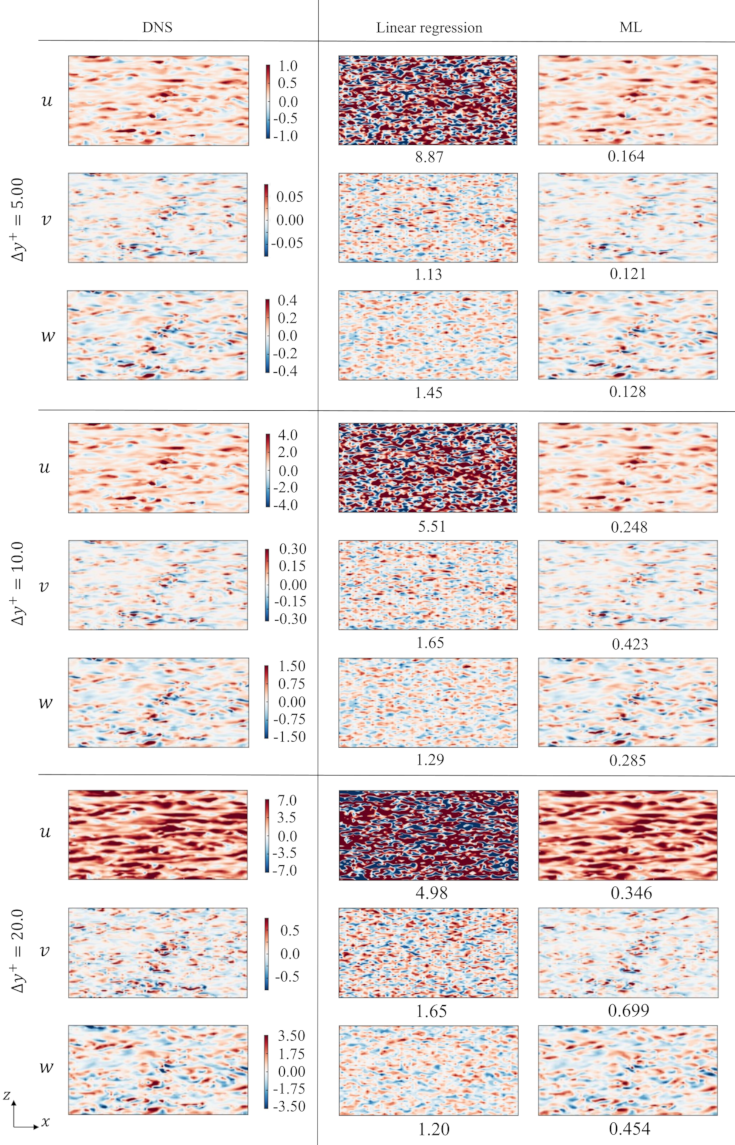}
  \caption{Visualization of virtual wall surface velocities for each grid-width case in {\it{a priori}} test. The values underneath each contour are the $L_2$ error norm.}
  \label{visualization}
\end{figure}
%%%%%%%%%%%%%%

%%%%%%%%%%%%%%
%  figure   %
%%%%%%%%%%%%%%
\begin{figure}
  \centering
  \includegraphics[width=1.\textwidth]{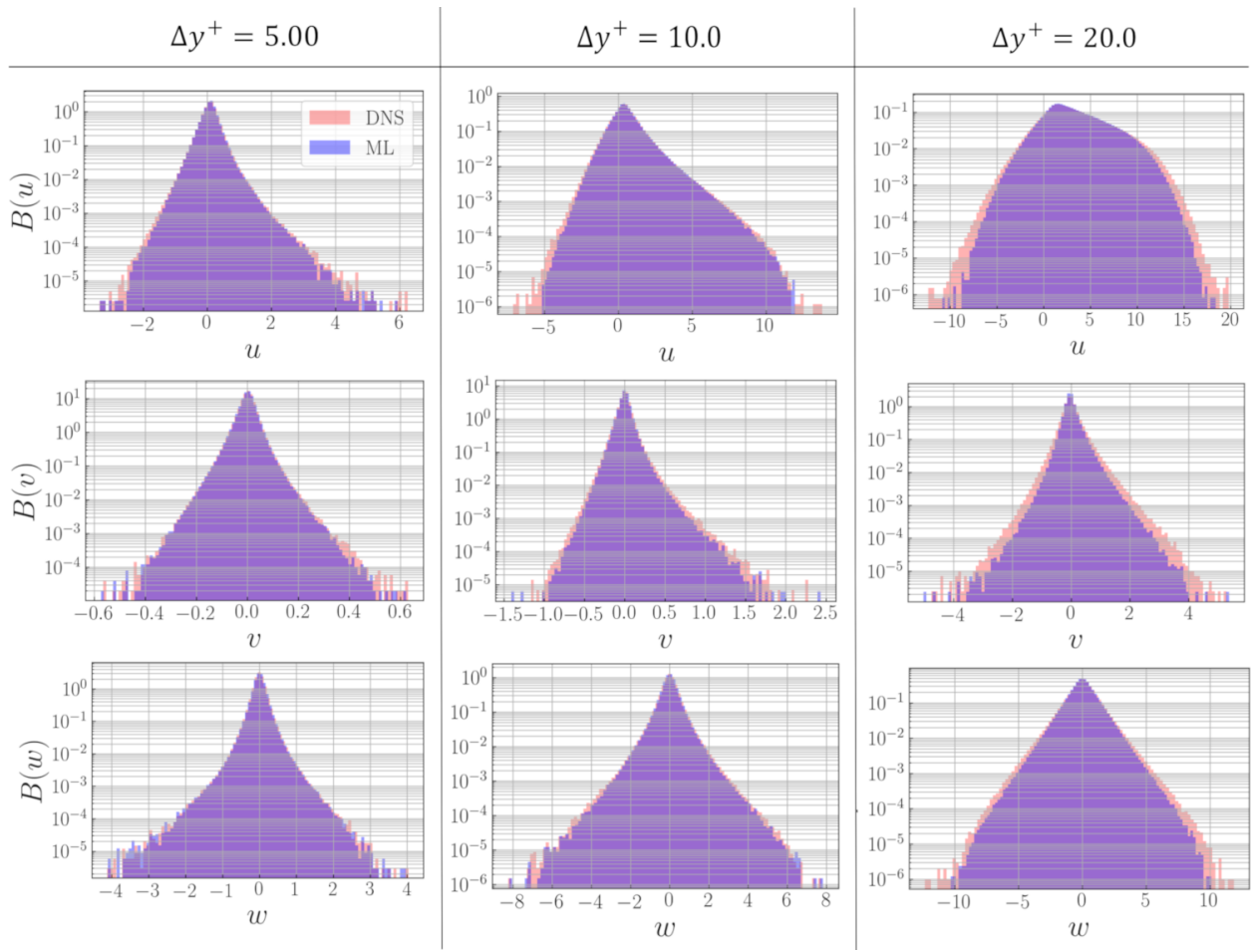}
  \caption{Probability density function of virtual wall surface velocity obtained by the DNS and machine-learning model in {\it{a priori}} test.}
  \label{apriori_PDF}
\end{figure}
%%%%%%%%%%%%%%

%%%%%%%%%%%%%%
%  figure   %
%%%%%%%%%%%%%%
\begin{figure}
  \centering
  \includegraphics[width=1.\textwidth]{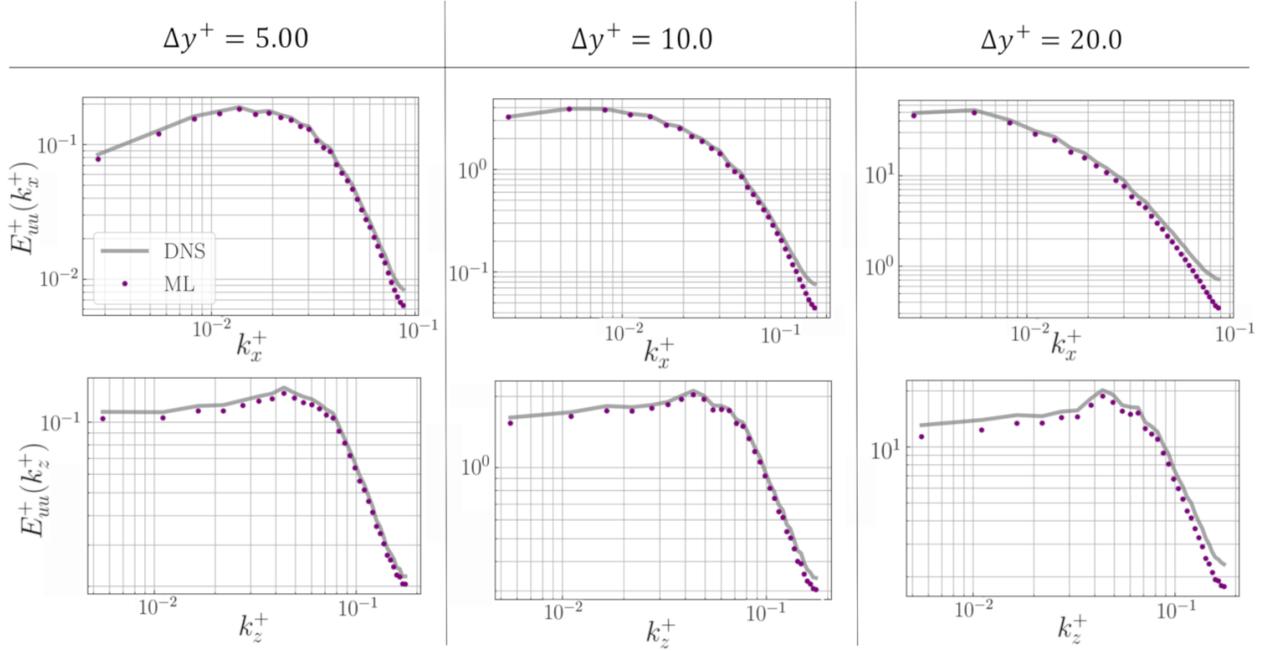}
  \caption{Kinetic energy spectrum in the streamwise and spanwise directions obtained by the DNS and machine-learning model in {\it{a priori} test}.}
  \label{apriori_enespe}
\end{figure}
%%%%%%%%%%%%%%

%%%%%%%%%%%%%%
%  figure   %
%%%%%%%%%%%%%%
\begin{figure}
  \centering
  \includegraphics[width=1.\textwidth]{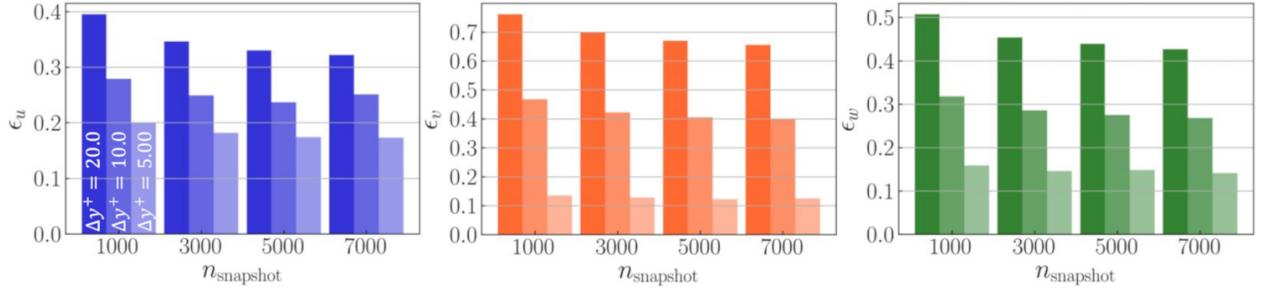}
  \caption{Dependence of the estimation accuracy on the number of training snapshots. The $L_2$ error norm for each velocity attribute is compared for each grid width.}
  \label{snap}
\end{figure}
%%%%%%%%%%%%%%

\section{Results}
\label{sec:3}

We apply the present technique to a turbulent channel flow for easiness of assessments.
As explained in Section~\ref{sec:2-1}, we use the DNS data for the training of the machine-learning model in {\it{a priori}} test.
The governing equations are the incompressible continuity equation and Navier--Stokes equation, 
\begin{align}
    \bm{\nabla}\cdot\bm{u}=0, ~~~ \frac{\partial \bm{u}}{\partial t}=-\bm{\nabla}\cdot (\bm{uu})-\bm{\nabla} p+\frac{1}{{\rm Re}_\tau}\nabla^{2}\bm{u}, 
    \label{eq_DNS}
\end{align}
where ${\bm u}=[u, v, w]^T$ represents the velocity vector in the streamwise $(x)$, wall-normal $(y)$ and spanwise $(z)$ directions; $p$ is the pressure and $t$ is the time. 
All physical quantities are made dimensionless by using density $\rho^{*}$, friction velocity $u_\tau^{*}$, and channel half-width $\delta^{*}$, where $(\cdot)^{*}$ denotes the dimensional quantities. 
The DNS is performed under the constant pressure gradient condition at the friction Reynolds number ${\rm Re}_\tau =(u_\tau ^{*}\delta^{*})/\nu^{*}=180$, where $\nu^{*}$ denotes the kinematic viscosity.
The size of computational domain and grid points here are $(L_x\times L_y\times L_z)=(4\pi\delta\times2\delta\times2\pi\delta)$ and $(N_x\times N_y\times N_z)=(256\times96\times256)$. 
The time step in the present DNS is $\Delta t^{+}_{\rm DNS}=6.30\times10^{-2}$. 
The present DNS code is the same as that used in the previous study \cite{nakamura2020extension}.
The governing equations~(Eq.~\ref{eq_DNS}) are spatially discretized with the energy-conserving fourth-order finite difference scheme on a staggered grid system~\cite{MLVM1998}. 
The temporal integration is performed using the low-storage, third-order Runge-Kutta/Crank–Nicolson scheme~\cite{SMR1991} with the higher-order SMAC-like velocity-pressure coupling scheme~\cite{DD1992}. 
The pressure Poisson equation is solved with the fast Fourier transform in the $x$ and $z$ directions and the tridiagonal matrix algorithm in the $y$ direction. 
No-slip boundary condition is imposed in the wall-normal direction and the periodic boundary condition is applied in the $x$ and $z$ directions.

In this study, three cases in terms of grid width in the wall-normal direction are considered, as summarized in Table~\ref{grid_width}. 
For the training of machine-learning model, we use the $x-z$ cross-sectional velocity data subsampled to $(64\times64)$, which are obtained by a direct numerical simulation (DNS) at the same friction Reynolds number ${\rm Re}_{\tau}= 180$ as that used in the baseline LES, as explained in Section~\ref{sec:2-1}. 
The time interval for the training data sampling is $\Delta t_{\rm ML}^+=1.26$, which corresponds to 20 time steps in the DNS. 
We use 3000 snapshots for training the baseline model, although we will discuss the dependence of the estimation ability on the amount of the training snapshots later.  
Among them, $70\%$ is used for the training data, while $30\%$ is used for the validation data.

\subsection{{\it A priori} test: construction of machine-learning-based wall model}

We construct a machine-learning model to estimate the virtual wall surface velocity, which will be applied for LES.
Let us visualize the estimated virtual wall surface velocities for each case in Fig.~\ref{visualization}.
The values underneath each contour represent the $L_2$ error norm normalized by the velocity fluctuation $\epsilon = {||\tilde{{\bm{u}}}_{\rm{DNS}}-\tilde{{\bm{u}}}_{\rm{ML}}||_2}/{||\tilde{{\bm {u}^\prime}}_{\rm{DNS}}||_2}$.
The linear regression cannot estimate the virtual wall surface velocity accurately in terms of both the contours and the $L_2$ error norm. 
In contrast, the flow fields estimated by the machine-learning model are in qualitative agreement with the reference DNS for all three cases.

We also investigate the ability of models using probability density function (PDF), as presented in Fig.~\ref{apriori_PDF}$(a)$.
The velocity distributions of DNS and machine-learning-based estimation are generally consistent, but the low probability events do not show good agreement with each other.
This is because the present model is trained to minimize the $L_2$ error as stated in Eq.~\ref{eq2}, thereby leading to output the average value of fluctuation components.

We then evaluate the estimation performance of the machine-learning model on wave space using the energy spectrum, as presented in Fig.~\ref{apriori_enespe}$(b)$.
For each grid width, the machine-learning model is able to estimate well especially the low wavenumber components.
However, the overestimation can be found at the high wavenumber counterparts, which implies that machine-learning model preferentially estimates the low wave-number components. 
This observation is consistent with previous studies of turbulence analysis using supervised machine learning methods~\cite{FFT2020b,SSS2020}.

The dependence of the estimation ability on the amount of the training snapshots is also examined in Fig.~\ref{snap}.
The estimation accuracy improves with increasing the number of snapshots used for training of machine-learning model in all cases.
Notably, the decreasing rate of the error becomes smaller with the cases of more than 3000 snapshots.
Therefore, we hereafter use machine-learning models trained with 3000 snapshots in {\it{a posteriori}} test.

%%%%%%%%%%%%%%%%%%%%%%%%%%%%%%%%%%%%%%%%%%%%%%%%%%%%%%%%%%%%%%%%
%%%%%%%%%%%%%%%%%%%%%%%%%%%%%%%%%%%%%%%%%%%%%%%%%%%%%%%%%%%%%%%%
%                 A posteriori test 
%%%%%%%%%%%%%%%%%%%%%%%%%%%%%%%%%%%%%%%%%%%%%%%%%%%%%%%%%%%%%%%%
%%%%%%%%%%%%%%%%%%%%%%%%%%%%%%%%%%%%%%%%%%%%%%%%%%%%%%%%%%%%%%%%

\subsection{{\it A posteriori} test: application of machine-learned model to LES}

In {\it{a posteriori}} test, the machine-learning model trained in {\it{a priori}} test is applied to the LES.
We use {\sffamily{f2py}}~\cite{F2PY} to combine a FORTRAN-based simulation codes with a python-based machine-learning module.
In LES, the governing equations are the filtered and coarse-grained continuity equation and the Navier--Stokes equation, 
\begin{align}
    \bm{\nabla}\cdot\bar{\bm{u}}=0, ~~~ \frac{\partial \bar{\bm{u}}}{\partial t} = - \bm{\nabla}\cdot({\bar{\bm{u}}}{\bar{\bm{u}}})-\bm{\nabla}\bar p+\frac{1}{{\rm Re}_\tau} \nabla^{2}\bar{\bm{u}}+\bm{\nabla}\cdot\bar{\bm\tau},
\end{align}
where $(\bar{\;\cdot\;})$ represents a filter operation, and $\bar{\bm\tau}$ denotes the sub-grid scale (SGS) stress tensor.
The size of the computational domain is the same as that of DNS, i.e., $(L_x\times L_y\times L_z) = (4\pi\delta\times2\delta\times2\pi\delta)$, and the number of grid points is $(N_x\times N_y\times N_z) = (64\times N^{\dag}_y\times64)$, where $N^{\dag}_y$ represents the number of grid points in the $y$ direction.
The uniform grid is used in the $y$ direction.
As the grid width in the $y$ direction, we consider three cases as summarized in Table~\ref{grid_width}. 
The time step in the present LES is $\Delta t^{+}=6.30\times10^{-2}$. 
We here use the constant Smagorinsky model~\cite{Smago1963} as the baseline SGS model. 
In the present demonstration, four cases of LES are compared as follows;
\begin{enumerate}
    \item LES without wall models (case 1),
    \item LES with van Driest's damping function~\cite{Driest1957} (case 2),
    \item LES assisted with machine-learned model, but without van Driest's damping function (case 3),
    \item LES assisted with machine-learned model and van Driest's damping function (case 4).
\end{enumerate}

%%%%%%%%%%%%%%
%  figure   %     
%%%%%%%%%%%%%%
\begin{figure}
  \centering
  \includegraphics[width=1.\textwidth]{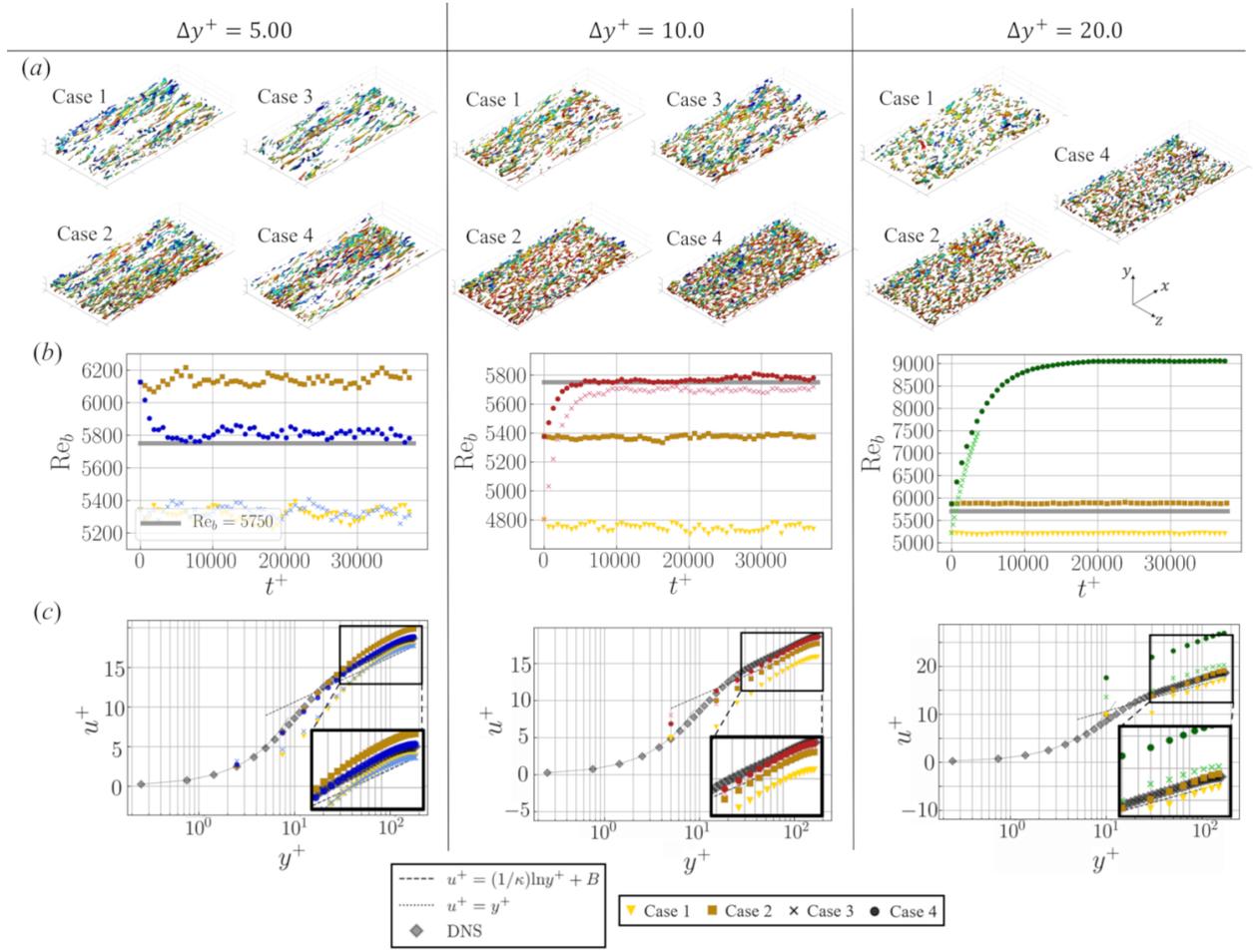}
  \caption{Summary of the present results in {\it a posteriori} test. $(a)$ the second invariant of the velocity gradient tensor $(Q^+=0.005)$, $(b)$ the time history of bulk Reynolds number ${\rm{Re}}_b$, and $(c)$ the mean streamwise velocity profile.}
  \label{Re_mean}
\end{figure}
%%%%%%%%%%%%%%

The flow fields obtained by each LES are visualized using the second invariant of the velocity gradient tensor $(Q^+=0.005)$~\cite{CPC1990} in Fig.~\ref{Re_mean}$(a)$. 
Note that we only compare among cases 1, 2, and 4 with $\Delta y^+=20.0$ because case 3 has shown an unstable behavior of the simulation due to the low accuracy of machine-learned model trained in {\it{a priori}} test.
These visualized fields exhibit the vortex structures in a reasonable manner.
The time history of bulk Reynolds number ${\rm{Re}}_b$ is also evaluated to investigate the correction of the simulation itself as shown in Fig.~\ref{Re_mean}$(b)$.
Note that the time history of case 3 with $\Delta y^+=20.0$ is only shown until around $t^+=5.00\times10^3$ due to the unstable simulation as mentioned above. 
The averaged bulk Reynolds numbers provided by DNS are also shown as the gray line in each case for comparison.
For each grid width, the bulk Reynolds number is not sufficiently corrected with case 3. 
On the other hand, case 4 is able to correct it with $\Delta y^+=\{5.00, 10.0\}$. 
Note that such correction cannot be observed with $\Delta y^+=20.0$. 
This is likely caused by the low estimation ability of the model trained in {\it{a priori}} test. 
Moreover, mean streamwise velocity profiles are also compared in Fig.~\ref{Re_mean}$(c)$.
Analogous to the observation in Fig.~\ref{Re_mean}$(b)$, case 4 shows its reasonable ability with $\Delta y^+=\{5.00, 10.0\}$. 
Note again that the overestimation with $\Delta y^+=20.0$ is likely caused by the lack of estimation ability as stated above. 
Hence, a reasonable performance of {\it{a priori}} test, at least, is required for the present correction method.

%%%%%%%%%%%%%%
%  figure   %     
%%%%%%%%%%%%%%
\begin{figure}
  \centering
  \includegraphics[width=0.92\textwidth]{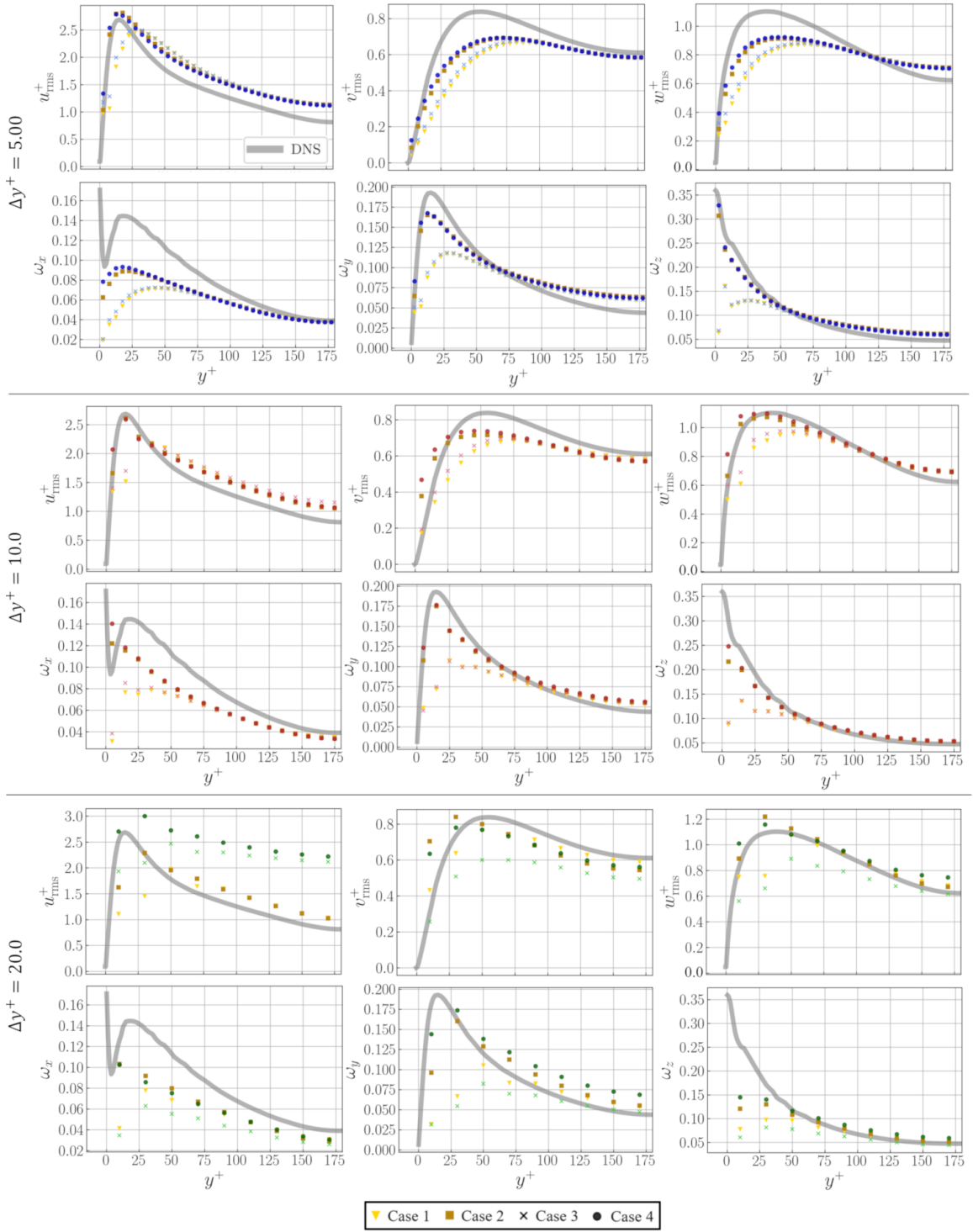}
  \caption{Root-mean squared values of velocity and vorticity fluctuation in {\it{a posteriori}} test.}
  \label{RMS_aposteriori}
\end{figure}
%%%%%%%%%%%%%%

To further examine the physical validity of the present LES in each case, the root-mean square (RMS) of velocity and vorticity fluctuations are summarized in Fig.~\ref{RMS_aposteriori}. 
The results in case 4 with $\Delta y^+=\{5.00, 10.0\}$ show closer distributions to that of the DNS compared to the other cases especially near the wall.
This is likely because the SGS viscosity is corrected by applying the van Driest's damping function, thereby leading to the correction of the RMS values near the wall.
Therefore, the utilization of both the van Driest model (i.e., the physical correction for the SGS model) and the machine-learned model (i.e., the correction for the error due to discretization of boundary condition) employ well to capture near wall behavior correctly.

\subsection{Influence on the used SGS model in LES}

%%%%%%%%%%%%%%
%  figure   %    
%%%%%%%%%%%%%%
\begin{figure}
  \centering
  \includegraphics[width=1.\textwidth]{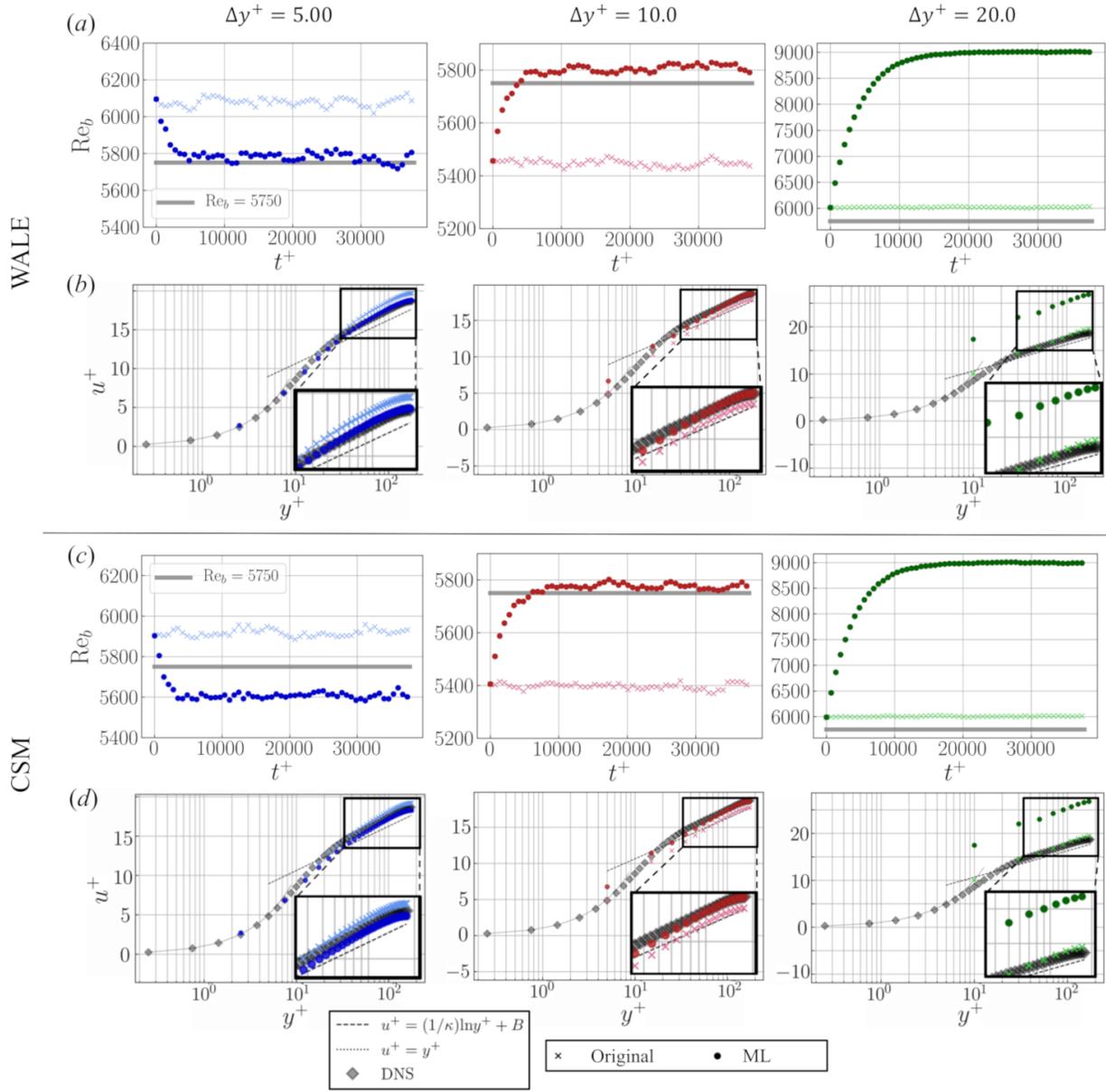}
  \caption{Comparison between the original LES and the LES assisted with machine-learning model (ML) using WALE model and CSM model. $(a)$ and $(c)$, the time history of bulk Reynolds number ${\rm{Re}}_b$; $(b)$ and $(d)$, the mean streamwise velocity profile. }
  \label{Re_mean_WALE_CSM}
\end{figure}
%%%%%%%%%%%%%%

As mentioned above, we have used the Smagorinsky model for the present analyses in {\it a posteriori} test. 
We here discuss the generalizability of the proposed machine-learning-based wall model with regard to SGS models. 
In addition to Smagorinsky model, let us consider two SGS models; Wall-Adapting Local Eddy-viscosity model (WALE)~\cite{ND1999}, and Coherent Structure Model (CSM)~\cite{Kobayashi2005}.
The results of time history of bulk Reynolds number and the mean streamwise velocity profile are shown in Fig.~\ref{Re_mean_WALE_CSM}. 
The augmentation of LES can be seen with $\Delta y^+=\{5.00, 10.0\}$ while failing with $\Delta y^+=20.0$, which is the same trend as the case with the constant Smagorinsky model.
Therefore, the proposed machine-learning-based wall model is robust against the choice of SGS model.

\subsection{Robustness of machine learning model for Reynolds numbers}

Since our method relies on the training data provided by DNS as expressed in Eq.~\ref{eq2}, of particular interest here is its capability at higher Reynolds numbers than that in its training process. 
Let us apply the machine-learned model trained at ${\rm{Re}}_\tau=180$ to the LES at ${\rm{Re}}_\tau=360$.
The size of computational domain and grid points in the LES at ${\rm{Re}}_\tau=360$ are $(L_x\times L_y\times L_z) = (2\pi\delta\times2\delta\times\pi\delta), (N_x\times N_y\times N_z) = (64\times N^*_y\times64)$). 
The time step in the present simulation is $\Delta t^{+}=6.30\times10^{-2}$.
A non-uniform grid is applied in the $y$ direction at $y^+>40$ multiplied by a given stretch rate, while the uniform grid is considered at $y^+<40$. 
%%%%%%%%%%%%%%%%%%%%%%%%%%%%%%%%%%%%%%%%%
%             table 2
%%%%%%%%%%%%%%%%%%%%%%%%%%%%%%%%%%%%%%%%%
\begin{table*}
    \vspace{3mm}
    \centering
    \vspace{3mm}
    \caption{The covered grid width in the $y$ direction at ${\rm{Re}}_\tau=360$ for {\it a posteriori} test. }
    \label{grid_width_360}
    \begin{tabular}{cccc}\hline
    \hline
        $\Delta y^+$ & Stretch rate & $\Delta y^+_{\rm{max}}$ & $N^*_y$\\
        \hline
        %\hline
        5.00 & 1.0290 & 14.9 & 88\\
        %\hline
        10.0 & 1.0265 & 18.7 & 54\\
        %\hline
        20.0 & 1.0290 & 29.6 & 30\\
        \hline
        \hline
    \end{tabular}
\end{table*}
%%%%%%%%%%%%%%%%%%%%%%%%%%%%%%%%%%%%%%%%%
%             table 2
%%%%%%%%%%%%%%%%%%%%%%%%%%%%%%%%%%%%%%%%%
The number of grid points in the $y$ direction $N^*_y$ depends on the grid width, as summarized in Table~\ref{grid_width_360}.
We here use the constant Smagorinsky model as the SGS model.

%%%%%%%%%%%%%%
%  figure   %    
%%%%%%%%%%%%%%
\begin{figure}
  \centering
  \includegraphics[width=1.\textwidth]{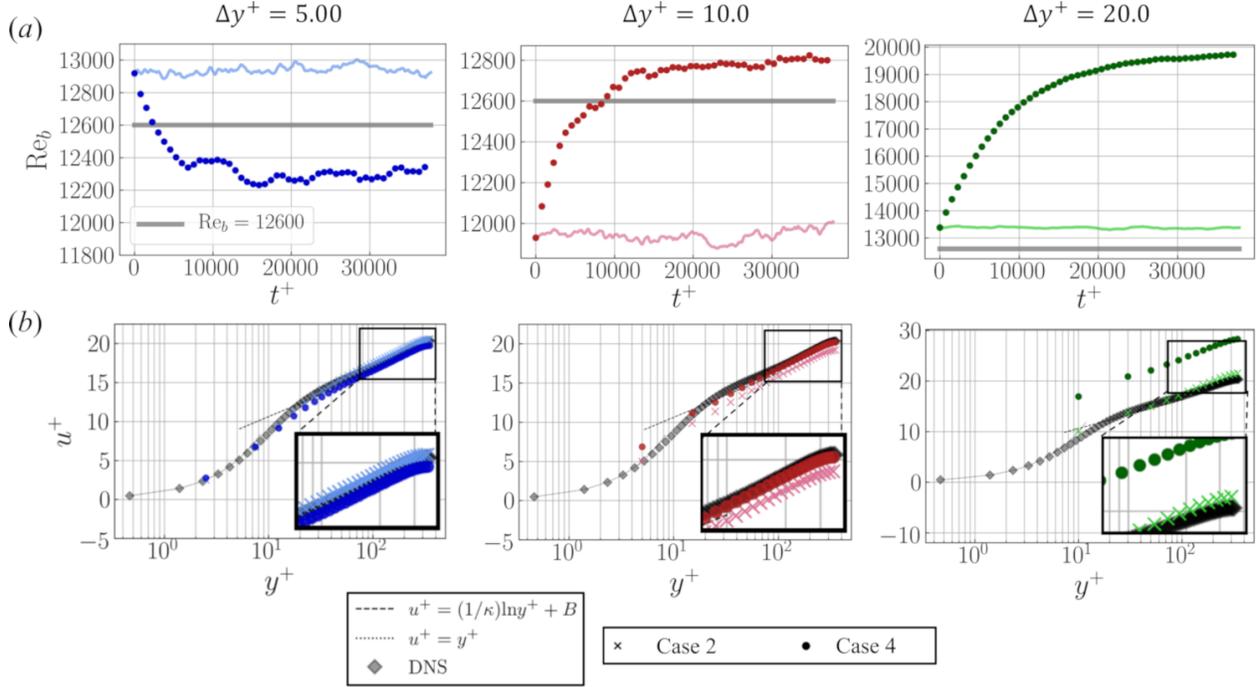}
  \caption{Comparison between case 2 (van Driest) and case 4 (ML trained at ${\rm{Re}}_\tau=180$ with van Driest) of the LES at ${\rm{Re}}_\tau=360$. $(a)$ Time history of bulk Reynolds number ${\rm{Re}}_b$ and $(b)$ mean streamwise velocity profile.}
  \label{Re_mean_360}
\end{figure}
%%%%%%%%%%%%%%

The performance of the present model (case 4) is assessed using the time history of bulk Reynolds number and the mean streamwise velocity profile in Fig.~\ref{Re_mean_360}.
For comparison, we also present the results of the DNS at ${\rm Re}_\tau = 360$ and case 2 which applies the van Driest function.
As can be expected, the model does not work with $\Delta y^+=20.0$ due to the lack of the estimation ability.
However, what is notable here is that the reasonable simulations can be achieved with $\Delta y^+=\{5.00, 10.0\}$ despite that the test Reynolds number is 
%in 
\fg{considered an}
extrapolation from the training range.
The present investigation suggests that we can expect a reasonable performance of a machine-learned model even at a higher Reynolds number for the grid width where the model employs well in a training Reynolds number range.

\subsection{Local learning for the present CNN model}

%%%%%%%%%%%%%%
%  figure   %    
%%%%%%%%%%%%%%
\begin{figure}
  \centering
  \includegraphics[width=1.\textwidth]{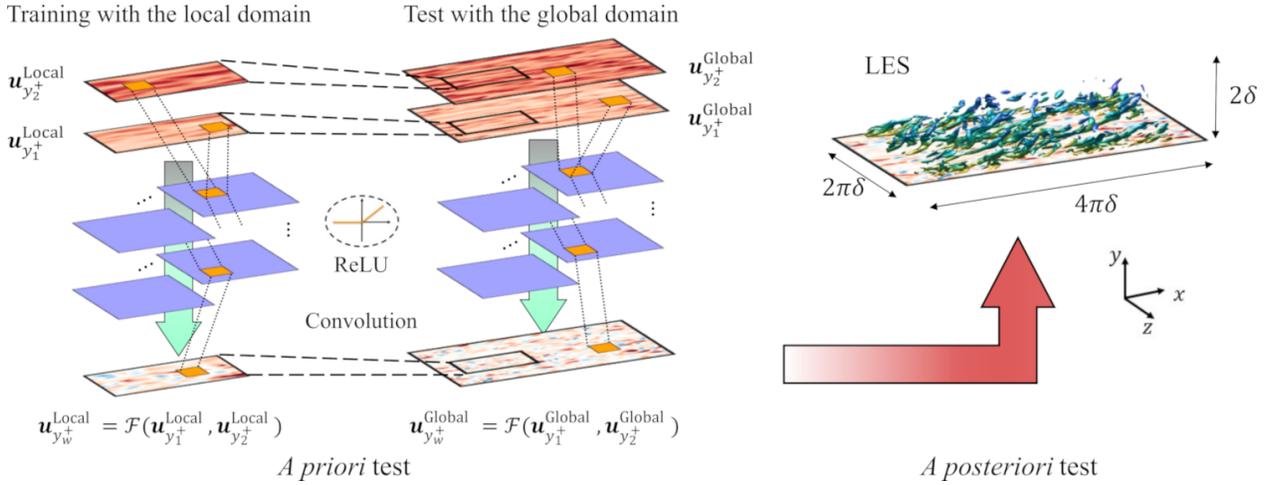}
  \caption{Application of locally trained ML model.}
  \label{Local_ML_model}
\end{figure}
%%%%%%%%%%%%%%

Towards practical applications of the proposed method, the generalizability of the machine learning model in terms of the domain size of the training data is preferable.
Hence, our interest here is the use of a locally trained ML model for both {\it a priori} and {\it a posteriori} tests.
One of the techniques for local domain training is to zoom-in and/or -out target images~\cite{MG2018}.
Morimoto et al.~\cite{morimoto2020generalization} has recently investigated the possibility of this concept and reported the effectiveness for various fluid flow data.

Let us use this zoom-in/out concept and discuss the dependence of the LES performance on the domain size in training data. 
We consider four cases in terms of the domain size of the training data; half $(L_x \times L_z) = (2\pi\delta\times\pi\delta)$, one quarter
$(\pi\delta\times\fg{(}1/2\fg{)}\pi\delta)$, and one eighth $(\fg{(}1/2\fg{)}\pi\delta\times\fg{(}1/4\fg{)}\pi\delta)$ compared to the global domain size.
The machine-learning model is trained with each local domain.
The constructed model is then applied to the global domain (i.e., $(L_x \times L_z) = (4\pi\delta\times2\pi\delta)$) in the test cases as summarized in Fig.~\ref{Local_ML_model}. 
We use 
%the baseline Reynolds number 
${\rm{Re}}_\tau=180$ as the training 
%data 
\fg{and test Reynolds number}
for this demonstration.

%%%%%%%%%%%%%%
%  figure   %    
%%%%%%%%%%%%%%
\begin{figure}
  \centering
  \includegraphics[width=0.97\textwidth]{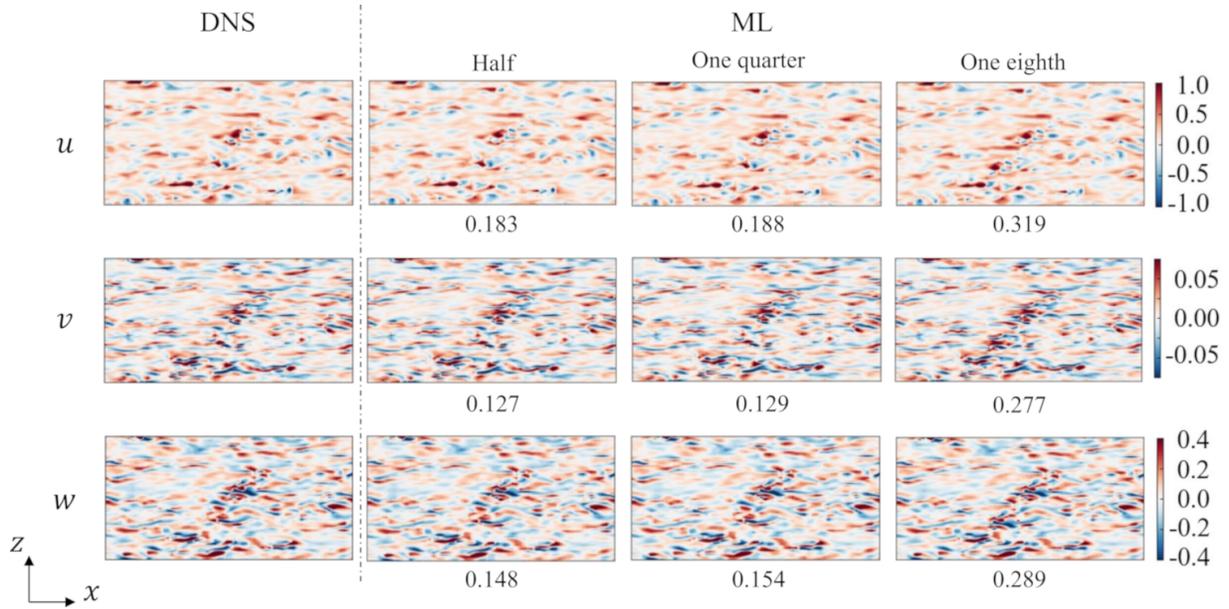}
  \caption{Virtual wall surface velocity estimated by the machine-learning model trained by each local domain with $\Delta y^+=5.00$ in {\it{a priori}} test. The values underneath each contour are the $L_2$ error norms.}
  \label{Local_apriori}
\end{figure}
%%%%%%%%%%%%%%

The virtual wall surface velocities with $\Delta y^+=5.00$ are shown in Fig.~\ref{Local_apriori}.
The flow fields estimated by the machine-learning models are in reasonable agreement with DNS data for all cases. 
On the other hand, the higher $L_2$ errors are shown with especially with the case of one eighth.
This is likely because the training data used as the input of the machine-learning model loses the low-wave number components by using zooming-in technique, which leads to decrease the estimation accuracy, since the present model dominantly estimates the low wavenumber components as discussed in Section 3.1.
%%%%%%%%%%%%%%
%  figure   %    
%%%%%%%%%%%%%%
\begin{figure}
  \centering
  \includegraphics[width=0.99\textwidth]{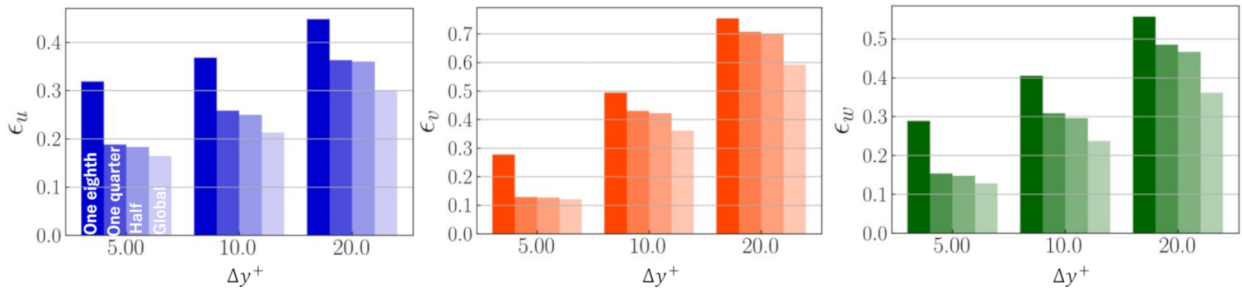}
  \caption{The $L_2$ error norms of locally trained cases in {\it a priori} test.}
  \label{L2_Local}
\end{figure}
%%%%%%%%%%%%%%
Moreover, the comparison of $L_2$ error in the other $y$ combinations are summarized in Fig.~\ref{L2_Local}. 
The similar trends can be found with $\Delta y^+=\{10.0, 20.0\}$ as well as with $\Delta y^+=5.00$.
Therefore, we hereafter use the machine-learned models trained with the domain size of one quarter in {\it{a posteriori}} test.

%%%%%%%%%%%%%%
%  figure   %    
%%%%%%%%%%%%%%
\begin{figure}
  \centering
  \includegraphics[width=0.95\textwidth]{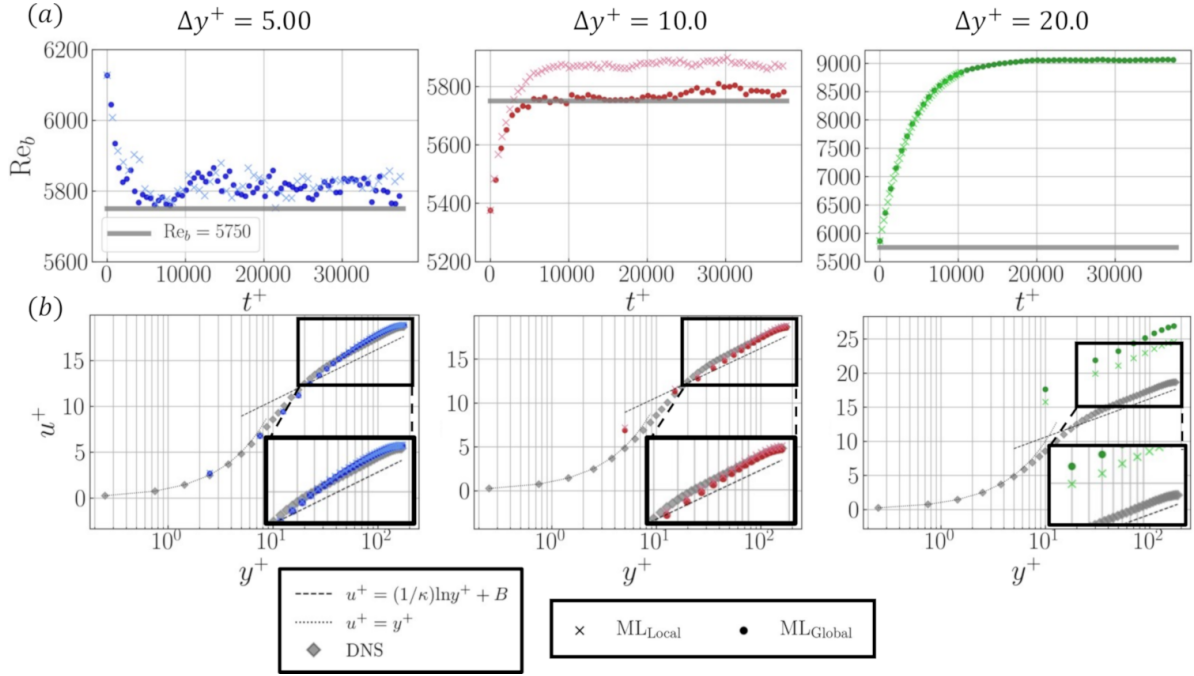}
  \caption{Comparison of the machine-learning-based wall model trained by the global domain and the local domain. $(a)$ the time history of bulk Reynolds number ${\rm{Re}}_b$ and $(b)$ the mean streamwise velocity profile.}
  \label{Re_mean_Local}
\end{figure}
%%%%%%%%%%%%%%

The results of time history of bulk Reynolds number and the mean streamwise velocity profile are shown in Fig.~\ref{Re_mean_Local}. 
Note that the time history with the machine-learning model trained by the local domain with $\Delta y^+=20.0$ is shown until around $t^+=1.00\times10^4$ due to the unstable simulation as the same reason mentioned in Section 3.2.
Although the fluctuations can be found with time history of bulk Reynolds number compared to the case with the global training, the LES can be augmented with $\Delta y^+=\{5.00, 10.0\}$, although not with $\Delta y^+=20.0$.
Summarizing above, the present model shows the generalizability in terms of the domain size used in a training pipeline.

\section{Conclusions}
\label{sec:4}

We assessed the performance of the machine-learning-assisted wall model for wall-resolved large-eddy simulation (LES) especially considering a coarse grid in the wall-normal direction.
In order to verify our idea, we first applied it to a turbulent channel flow at ${\rm Re}_{\tau}=180$.
In {\it{a priori}} test, we constructed the machine-learning models based on convolutional neural network (CNN) that estimate the artificial slip velocity from the velocity of two $x-z$ cross-sections in the region near the wall.
The constructed model was able to estimate the virtual wall-surface velocity well compared to the linear regression method. 
The machine-learned model was then applied to the present LES in {\it{a posteriori}} test. 
We found that the LES with a coarse wall-normal grid of $\Delta y^+=10$ can be augmented by the proposed model.

We further examined the dependence of the LES performance on the choice of SGS models and the Reynolds numbers. 
The robustness of the proposed model can be observed for both perspectives even if the cases are
%in 
extrapolation from the training range. 
The generalizability of the proposed model in terms of the domain size of the training data was finally investigated, which achieves the reasonable simulation performance compared to the cases with the global training.

We have several outlooks to improve the capability of the present CNN-based velocity estimator.
For example, we can consider the probabilistic neural network (PNN)~\cite{MFRFT2020} to quantify the uncertainty of its estimation.
This view is quite important in the present analysis where the accuracy of models trained in {\it a priori} test highly affects the performance in {\it a posteriori} test, since the PNN can tell us how we can rely on results provided by models.
Otherwise, unsupervised frameworks may also be helpful for the case where we have no solution data, as well discussed in Kim et al.~\cite{KKWL2021}.
Moreover, the combination with temporal prediction models, e.g., long short-term memory~\cite{nakamura2020extension,SGASV2019,eivazi2021recurrent,HFMF2019,HFMF2020a,HFMF2020b} and reservoir computing~\cite{pandey2020reservoir}, is also a considerable path for correcting the error due to temporal discretization 
when the present method is used in LES with a substantially larger computational time step.
Although the aforementioned extensions are just examples, we hope that the present paper is able to serve as a significant step to establish the data-driven LES wall model.

\section*{Acknowledgements}

This work was supported through JSPS KAKENHI Grant Number 18H03758 by Japan Society for the Promotion of Science.
We thank Mr. Mitsuaki Matsuo (Keio University) for fruitful discussions.

\section*{Declaration of interest}

The authors report no conflict of interest.

% \bibliographystyle{spmpsci}      % mathematics and physical sciences
% \bibliographystyle{spphys}       % APS-like style for physics
\bibliographystyle{unsrt}
%\bibliography{references}  % name your BibTeX data base

% \begin{thebibliography}{}
% %
% % and use \bibitem to create references. Consult the Instructions
% % for authors for reference list style.
% % Format for Journal Reference
% Author, Article title, Journal, Volume, page numbers (year)
% Author, Book title, page numbers. Publisher, place (year)
% \end{thebibliography}

\end{document}